# Finger Paint and Physics:
## A simple demonstration of Circular Motion and Conservation of Energy

Sarah Phan-Budd

*Department of Physics, Winona State University, Winona, MN, 55987*

It can be a challenge to come up with simple demonstrations of circular motion and conservation of energy. One such demonstration consists of a large exercise ball, off of which a small solid ball is rolled. The small ball is coated in finger paint so, after an initial push, it rolls nearly without slipping and creates a visible track.

Figure 1 shows the demonstration setup. A small solid metal ball is used for this experiment. The larger exercise ball is set in a container to keep it stable. The floor is covered with a tarp or other materials to protect from paint drips. Students are asked to model the length of the paint track.

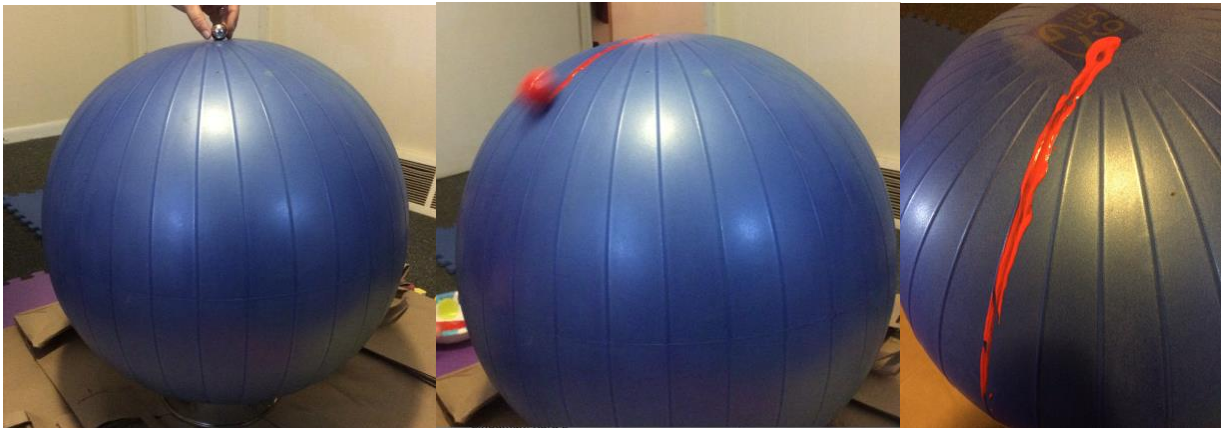

Figure 1: Demonstration Setup. The small solid ball is released from the top of the exercise ball, leaving a visible paint track which can be measured.

A similar problem is found in many intermediate level mechanics textbooks.[i] A small ball of mass m and radius r that has been released from rest at the top of a large sphere (radius R) rolls through a vertical distance h and an angle θ before losing contact with the large sphere, as shown in Figure 2.

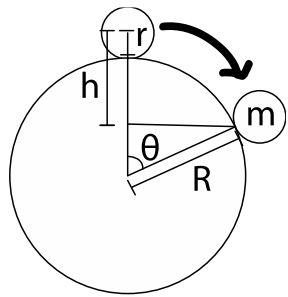

Figure 2: A small ball (mass m, radius r) rolls down a large sphere (radius R), falling a distance h through an angle θ.

We consider only the case where the small ball rolls entirely without slipping.
The following conservation of energy formula applies,
$$mgh = \frac{1}{2}mv^2 + \frac{1}{2}I\omega^2. \tag{1}$$

Here, $v = \omega r$ and $I = \frac{2}{5}mr^2$. As shown in Figure 1, $h = R + r - (R + r)\cos\theta$. Solving for the speed,
$$v^2 = \frac{10g(R+r)(1-\cos\theta)}{7}. \tag{2}$$

In the case that the ball rolls entirely without slipping, the radial equation of motion can also be applied, giving
$$mg\cos\theta - N = \frac{mv^2}{R+r}. \tag{3}$$
Here N is the normal force. The ball leaves the sphere when the normal force is zero, or
$$v^2 = g(R+r)\cos\theta. \tag{4}$$
Setting the speed obtained from the energy equation equal to the speed obtained from the equation of motion
$$\cos\theta = \frac{10}{17}. \tag{5}$$
This implies that the solid ball leaves the sphere at a release angle of about 0.94 radians or 54°. This is larger than the case of pure slipping, where frictionless object leaves the large ball when $\cos\theta = \frac{2}{3}$ at about 0.84 radians or 48°.

The circumference, C, of the exercise ball is given by $C = 2\pi R$. The length of the paint track, s, is given by the arclength formula $s = R\theta$ when the angle is measured in radians. Combining these two formulas, length of the paint track is given by.
$$s = \frac{C\theta}{2\pi}. \tag{6}$$
The lengths of the tracks and hence the release angle can be measured and compared with the no-slip and frictionless models. The demonstration was run 10 times on an exercise ball with a circumference of 180.9 cm. The average track length was $26.3 \pm 0.3$ cm (std. dev). This yields a release angle of $0.91 \pm 0.01$ rad or $52° \pm 1°$, demonstrating the non-slip model is a better fit. Data is given in Table 1.

| Track Length (cm) | Release Angle (radians) | Release Angle (degrees) |
| --- | --- | --- |
| 26.6 | 0.92 | 53 |
| 26.2 | 0.91 | 52 |
| 26.4 | 0.92 | 53 |
| 27.0 | 0.94 | 54 |
| 27.0 | 0.94 | 54 |
| 24.2 | 0.84 | 48 |
| 26.6 | 0.92 | 53 |
| 26.0 | 0.90 | 52 |
| 27.0 | 0.94 | 54 |
| 26.0 | 0.90 | 52 |

Table 1: Track length measurements and calculated release angle measurements for a small solid ball rolling off a large exercise ball.

The tracks tend to be a little shorter than predicted by the no-slip model. This is consistent with what one would expect with a more realistic model where the ball slips before falling off the large sphere. [ii] [iii] In fact, the tracks should be even a bit shorter than the rolling plus slipping model, since slipping objects do not leave paint tracks for the entire duration of their slip. It is worthwhile to have a classroom discussion of sources of deviation from our simple model, and there are a number of excellent papers analyzing the motion of a sphere slipping [iv] [v][vi], rolling on a grooved track[vii] [viii]and rolling and slipping on a grooved track. [ix]

A final word to the wise. As your local pre-school teacher will tell you, putting just a little bit of soap in the finger paint helps it wash off more easily.

---

[i] Herbert Goldstein, Charles Poole and John Safko, *Classical Mechanics,* 3rd ed. (Addison-Wesley, 2002), p. 66.

[ii] J. Flores, A.G. del Rio, A. Calles and H. Riveros, "A Simple Problem In Mechanics: A Qualitative Approach", *Am. J. of Phys*. **40**, 595 (1972).

[iii] V. Jayanth, C. Raghunadan and A. Biswas, "A sphere moving down the surface of a static sphere and a simple phase diagram" https://arxiv.org/abs/0808.3531

[iv] Tom Prior and E.J. Mele, "A block slipping on a sphere with friction: Exact and perturbative solutions", *Am. J. of Phys*. **75**, 423 (2007).

[v] O. L. de Lange, J. Pierrus, Tom Prior and E. J. Mele, "Comment on "A block slipping on a sphere with friction: Exact and perturbative solutions*," Am. J. of Phys*. **76,** 92 (2008).

[vi] C. E. Mungan, "Sliding on the Surface of a Rough Sphere", *Phys. Teach.* **41**, 326 (2003).

[vii] R. A. Bachman, "Sphere rolling down a grooved track", *Am. J. of Phys*. **53**, 765(1985).

[viii] Qing-gong Song, "The requirement of a sphere rolling without slipping down a grooved track for the coefficient of static friction", *Am. J. of Phys*. **56,** 1145( 1988).

[ix] D. C. de Souza and V. R. Coluci, "The Motion of a Ball Moving Down a Circular Path", *Am. J. of Phys*. **85**, 124 (2017).